\documentclass[11pt,preprint]{aastex}

\newcommand{\etal}{{\it et~al.}}

\usepackage{longtable}

\begin{document}

\title{Small and Nearby NEOs Observed by NEOWISE During the First Three Years of Survey: Physical Properties}

\author{Joseph R. Masiero\altaffilmark{1}, E. Redwing\altaffilmark{1,2}, A.K. Mainzer\altaffilmark{1}, J.M. Bauer\altaffilmark{3}, R.M. Cutri\altaffilmark{4}, T. Grav\altaffilmark{5}, E. Kramer\altaffilmark{1}, C.R. Nugent\altaffilmark{4}, S. Sonnett\altaffilmark{5}, E.L. Wright\altaffilmark{6}}

\altaffiltext{1}{Jet Propulsion Laboratory/California Institute of Technology, 4800 Oak Grove Dr., MS 183-301, Pasadena, CA 91109, USA, {\it Joseph.Masiero@jpl.nasa.gov}}
\altaffiltext{2}{The Pennsylvania State University, State College, PA 16801, USA}
\altaffiltext{3}{University of Maryland, College Park, MD 20742, USA}
\altaffiltext{4}{California Institute of Technology, IPAC, 1200 California Blvd, Pasadena, CA 91125 USA}
\altaffiltext{5}{Planetary Science Institute, Tucson, AZ 85719 USA}
\altaffiltext{6}{University of California, Los Angeles, CA, 90095}

\begin{abstract}

Automated asteroid detection routines set requirements on the number
of detections, signal-to-noise ratio, and the linearity of the
expected motion in order to balance completeness, reliability, and
time delay after data acquisition when identifying moving object
tracklets.  However, when the full-frame data from a survey are
archived, they can be searched later for asteroids that were below the
initial detection thresholds. We have conducted such a search of the
first three years of the reactivated NEOWISE data, looking for
near-Earth objects discovered by ground-based surveys that have
previously unreported thermal infrared data.  Using these
measurements, we can then perform thermal modeling to measure the
diameters and albedos of these objects.  We present new physical
properties for $116$ Near-Earth Objects found in this search.

\end{abstract}

\section{Introduction}

The Near-Earth Object Wide-field Infrared Survey Explorer (NEOWISE)
has been surveying the sky at two thermal infrared wavelengths since
2013 December 13 \citep{mainzer14neowise}.  As part of regular
operations, NEOWISE uses the WISE Moving Object Processing System
(WMOPS) to identify transient sources and link them into tracklets
that consist of $5$ observations or more
\citep{mainzer11,cutri15}.  This length limit was set to balance the
reliability of candidate moving object tracklets with survey
completeness, while fulfilling the mission requirement of reporting
moving object detections to the Minor Planet Center within $10$ days
of the tracklet midpoint.

There are cases where a near-Earth object (NEO) was bright enough to
be detected by NEOWISE, but observed an insufficient number of times
to be registered by WMOPS.  Frequently these are objects that are
passing very close to the Earth near the time of discovery, and thus
move through the NEOWISE field of regard very rapidly.  In other
instances, an object's motion may be changing over the observations
beyond the set tolerance of the WMOPS software.  Because NEOWISE
archives and makes available all single-exposure observations, it is
possible to search these images for NEO detections missed by the
automated pipeline.

In this work, we present such a search of the first three years of the
reactivated NEOWISE survey data for close-pass NEOs, similar to the
one carried out for the cryogenic NEOWISE data by
\citet{mainzer14tinyneo}.  The goal of this search is to increase the
number of NEOs with physical characterization, and lengthen the orbital
arcs of short-arc NEOs with incidental observations beyond the
timescale of archived ground-based observations.

\section{Methods}

Small NEOs are preferentially discovered when they are close to the
Earth, as they are at their brightest and thus are easiest to detect.
We searched the NEOWISE images for objects in the Minor Planet
Center's list of NEO orbital elements\footnote{\it
  https://www.minorplanetcenter.net}, focusing on objects with
provisional designations indicating that they had been discovered in
2014, 2015, or 2016.  Although it is possible that some numbered or
multi-opposition NEOs were relatively nearby Earth during those years,
the astrometry provided by recovered NEOWISE detections will not
significantly improve the orbit, and later automated recovery will be
easier for these objects.  Thus, we focus this work on objects where
the observations may provide the biggest gain.  Future work will
include a more comprehensive search for all known objects that are at,
or just below, the single-exposure detection limits.

We used the Solar System Object search tool provided by the NASA/IPAC
Infrared Science Archive (IRSA) in the WISE image server\footnote{\it
  https://irsa.ipac.caltech.edu/applications/wise/} to determine the
predicted locations for all NEOs with provisional designations
assigned from the start of 2014 to the end of 2016 for which
detections had not already been reported by NEOWISE.  This tool uses
the spacecraft position as well as the propagated orbit of the
asteroid from JPL Horizons\footnote{\it
  https://ssd.jpl.nasa.gov/horizons.cgi} to determine if the object
was coincident with any recorded NEOWISE image at the time that
image was acquired.  Our search included both bands acquired by the
NEOWISE survey: $3.4 \mu$m (W1) and $4.6 \mu$m (W2).

While we generally restricted our search to objects with small
positional uncertainties at the time of the NEOWISE observations, a
broader search for high SNR detections for objects with larger
uncertainties was also carried out.  For $51$ objects, the search of
the NEOWISE archive recovered observations that fell in between
previously published observations and thus had positional
uncertainties smaller than the size of the point spread function
(PSF), approximately 6.5 arcsec.  The remaining $71$ objects
were detected by NEOWISE outside of the published observation arc. Of
these, three had other observations obtained or linked subsequent to
the submission of our astrometry (including 2016 UH$_{101}$ which was
linked to 2010 PP$_{58}$, an NEO originally discovered by WISE during
its cryogenic survey).  For $44$ of the objects where the NEOWISE
detections extended the observational arc, multiple NEOWISE detections
were obtained along the track at the same positional offsets,
confirming these associations as correct.

The remaining $24$ objects were observed only a single time, and thus
are the most difficult cases to confirm the association between the
detection and the object.  Five of the single-detections were within a
$10~$arcsec search radius of the predicted position and had a
SNR$_{W2}>5$, making a positive association of the source with the
object highly likely.  A wider search revealed $15$ single detections
within $1~$arcmin of the predicted position and with SNR$_{W2}>10$ and
low PSF-fit residuals, making them unlikely to be noise.  The majority
of these also had SNR$_{W1}>3$, further strengthening the association
due to the fact that NEOs tend to be warm and are brighter in W2 than
W1.

The remaining $4$ sources (2015 LM$_{21}$, 2015 VU$_{65}$, 2016
TX$_{17}$ and 2016 XA$_{18}$) are lower-confidence associations, but
were included as they were deemed to be highly likely to be point
sources based on visual inspection.  The observations for both 2015
LM$_{21}$ and 2016 TX$_{17}$ were within 1 arcmin of the predicted
position, and had SNR$_{W1}>3$ and SNR$_{W2}>5$, leading us to
conclude these were likely correct associations.  The observation of
2016 XA$_{18}$ was within 1 arcmin of the predicted position and had
an SNR$_{W2}>5$, but only an SNR$_{W1}=2$.  This object was located in
the overlap region between survey exposures, and the image from
$11~$seconds later shows a source at the same position, though it was
too close to the image edge to be picked up by the pipeline source
extraction.  This second image supports our association between the
detection and object.  Finally, the associated observation with 2015
VU$_{65}$ was nearly 3 arcmin from the predicted position, however the
source was detected at SNR$_{W2}=20$ with a low residual for the fit
of the PSF to the detection, making it unlikely to be a cosmic ray.

Following the procedure used in our previous publications
\citep[e.g.][etc.]{mainzer14tinyneo,masiero17} we took all the
detections of our objects of interest that had been submitted to,
accepted by, and published by the Minor Planet Center (MPC), and
searched the NEOWISE Reactivation Database L1b source table on IRSA.
For each detection we search within 6 arcsec of the reported position,
with a further constraint of $5~$seconds on the difference between the
reported and observed MJD.  This query returned the 3.4~$\mu$m and
4.6~$\mu$m profile-fit magnitudes and associated errors for each
detection, as well as any coincident sources in the AllWISE Atlas
detection database.  The AllWISE Atlas is a deep-stack of the WISE
primary-mission data with the number of coadded images rising from
$\sim8$ at the ecliptic to $>100$ closer to the poles\citep{cutri14}.
This ensures that the detection in the NEOWISE survey data is not of a
faint background object with varying brightness.  We searched the
Atlas at the position of each detection with a radius of $6~$arcsec,
and rejected any detection that was found with $SNR>7$ in either band
in the Atlas. Because Atlas sources are extracted from stacks of $>10$
single-frame images, this SNR cut will capture high-confidence
background objects significantly below our nominal detection limit
that may be experiencing a transient brightening event, while
minimizing the chances of rejecting a detection due to a much fainter
coincident background source.

NEOWISE simultaneously acquires images at $3.4~\mu$m and $4.6~\mu$m
(referred to as W1 and W2, respectively).  For objects detected in
both bands, we require that the object has an NEO-like color
(i.e. $W1-W2>1~$mag, as opposed to stars that usually have a color of
$W1-W2\sim0$).  We detect $33$ asteroids in a single NEOWISE exposure
set and $89$ in multiple exposures, for a total of $354$
visually-confirmed detections of $122$ NEOs.  The astrometric
observations of these objects were reported to the Minor Planet
Center, and the orbital solution that was computed after the inclusion
of these observations was used for our thermal fitting to ensure the
most accurate distance measurements at the time of observation for
these close-pass objects.  We use the updated orbital elements, along
with the time of observation and reported spacecraft positions to
calculate the heliocentric and geocentric distances and phase angle at
the time of observation.  These parameters are also available via JPL
Horizons.

\begin{figure}[ht] 
\begin{center}
\includegraphics[scale=0.4]{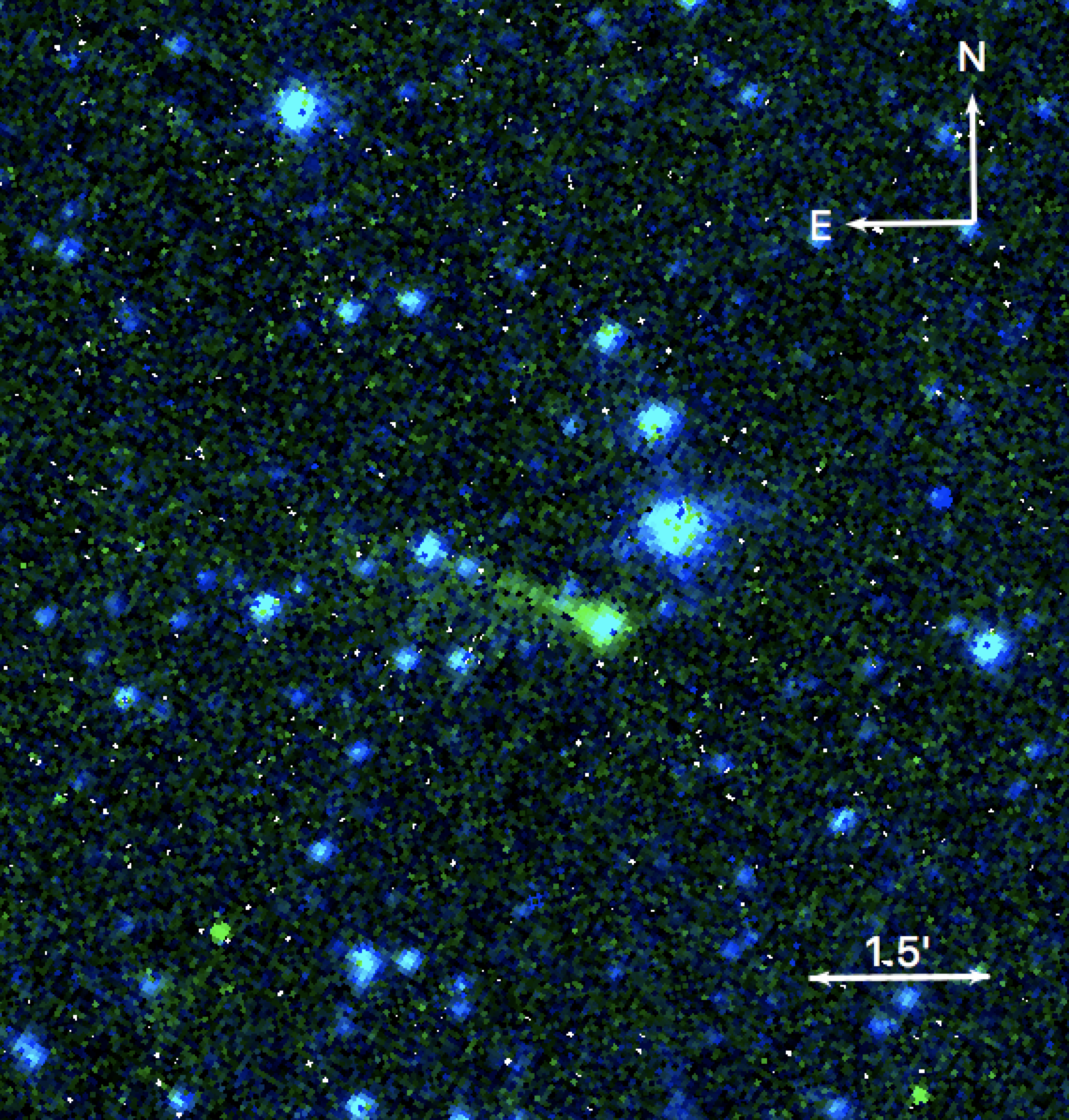}
\protect\caption{NEOWISE 2-band image of near-Earth object 2014 XK$_6$
  (center) showing clear signs of cometary activity with a tail
  extending $\sim1.5$ arcmin to the ENE, consistent with the anti-sunward
  direction.  W1 ($3.4~\mu$m) is shown in blue, and W2 ($4.6~\mu$m) is
  shown in green.}
\label{fig.comet}
\end{center}
\end{figure}

During the course of our search we identified one object, 2014 XK$_6$,
that showed clear signs of cometary activity.  This NEO was observed a
single time by NEOWISE, and detected in both bands.  We present the
2-band NEOWISE image in Figure~\ref{fig.comet}, which clearly shows
the cometary activity detected.  Astrometry for this object was
reported to the MPC, but no other detections of activity have been
found by other surveys, including in the Pan-STARRS discovery images
obtained one month after the NEOWISE detection (Weryk et al., 2017,
private communication).  2014 XK$_6$ has a Tisserand parameter of
$T_J=2.9$, consistent with a Jupiter-Family Comet.  It is possible
that this object is similar to near-Earth object (3552) Don Quixote,
which also revealed a tail in the infrared that was not seen in
optical measurements \citep{mommert14}, indicating either weak,
sporadic activity, a large contribution from CO$_2$ emission (which
falls in the W2 bandpass), or both.

\clearpage

We note that for $19$ of our targets, we find 5 or more detections in
the WISE single exposures.  In theory these nominally should have been
detected by WMOPS.  In $6$ of these cases, the object fell below the
single-exposure detection limit of SNR$=4.5$ used by WMOPS and thus
did not have sufficient data to construct a tracklet.  In $6$ other
cases, the object was observed near the ecliptic or equatorial poles,
where assumptions about the linearity of short-term on-sky motion in
the rectilinear coordinate system made by WMOPS are not valid.  For
the remaining $7$ objects, during close approach to Earth their
tracklets deviated from linear motion significantly enough over the
time span covered by the NEOWISE observations to not be linked by the
WMOPS software.  These objects all present compelling evidence for the
benefits of archiving all recorded images from surveys to allow for
later re-analysis.

\section{Thermal Modeling}

We use the Near-Earth Asteroid Thermal Model
\citep[NEATM,][]{harris98} to determine the physical properties of the
observed NEOs, following the process and selection criteria described
in our previous work \citep[e.g.][]{nugent15,nugent16,masiero17}.  In
brief, we extract NEOWISE W1 and W2 magnitudes and errors using IRSA
with the positions and times reported to the Minor Planet Center.  All
of our detections were visually inspected to ensure they were not
incorrect associations with cosmic rays, diffraction spikes,
nebulosity, or other sources of false detections, and over half of
our objects were detected in two bands at SNR$>3$.  Using the
MPC-published $H$ and $G$ photometric parameters for each object along
with the orbit, we determine the expected visible brightness at the
time of each observation.  As most of our observations tend to be
close in time to ground-based detections, this method should in
general result in reasonable estimates for the visible brightness even
if the $H$ and $G$ parameters are not well-constrained.  However,
systematic errors induced by the assumed color corrections known to be
present in the published MPC $H$ magnitudes will still impact our fits
\citep[cf.][]{veres15,masiero17}.

Included in the assumptions we make for our thermal modeling is that
the emissivities of the asteroids are uniformly $\epsilon=0.9$ at all
bands.  As pointed out by \citet{mhyrvold18}, this may violate
Kirchhoff's law if the point reflectance of the surface material is
not $0.1$. However, as emissivity and beaming parameter jointly modify
the characteristic temperature modeled by NEATM, the uncertainty in
emissivity (typically of order $10\%$) is subsumed by the much larger
uncertainty on the beaming parameter (of order $50\%$ for NEOs).  In
fact, some of the variation in beaming parameter is likely due to
variations in emissivity, and thus these variations are properly
accounted for in our model.

\begin{figure}[ht] 
\begin{center}
\includegraphics[scale=0.6]{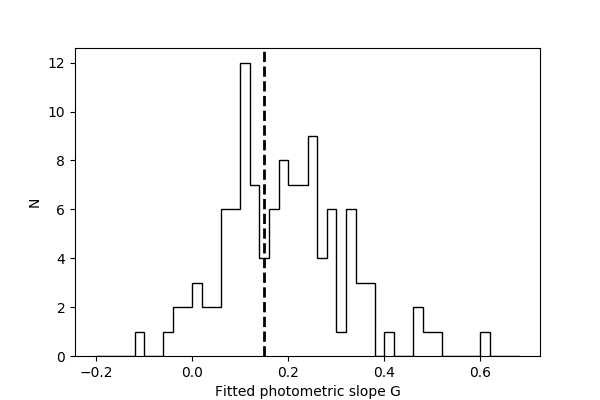}
\protect\caption{Distribution of all fitted photometric $G$ slopes in
  the \citet{tholen09} PDS archive.  The default assumed value most
  commonly used for asteroids, $G=0.15$, is shown as a vertical dashed
  line.  Fitted values can vary by over $100\%$ from the assumed
  value.}
\label{fig.Gslope}
\end{center}
\end{figure}

Further, \citet{mhyrvold18} shows that the vast majority of asteroids
and analogs have emissivities within $\sim10\%$ of $0.9$.  An
important caveat is that \citet{mhyrvold18} neglected to account for
the uncertainty on the $G$ phase slope parameter when calculating
emissivity from measured albedos; the true range of variations of $G$
away from the default assumption of $0.15$ \citep[of order $66\%$,
  cf.][]{lagerkvist90}, when included in the calculation, results in
an uncertainty on emissivity that means the determined values are
consistent with $0.9$ for nearly all cases, further validating our
assumptions.  As an example of the variation possible in the $G$
parameter, we show in Figure~\ref{fig.Gslope} all values of $G$ listed
as fitted in the PDS Asteroid Absolute Magnitude and Slope dataset
\citep{tholen09}.  This histogram shows that the default assumed value
of $G=0.15$, while being roughly a median value for the population,
falls between the peaks of the weakly bimodal distribution of $G$
values.  Additionally, fitted $G$ parameters cover a large range of
values, which would propagate directly to a large uncertainty when
calculating emissivity from $G$.  The relation between emissivity and
geometric albedo \citep[from][]{bowell89} is written:

  \[1- \epsilon = A = q p_V = (0.29 + 0.684 G) p_V\]

The effect of the possible range of values of $G$ on the final
computed $\epsilon$ depends on the albedo of the material, with higher
albedo materials showing a more significant effect.  However, the
uncertainty on $G$ cannot be ignored categorically.

\begin{figure}[ht] 
\begin{center}
\includegraphics[scale=0.6]{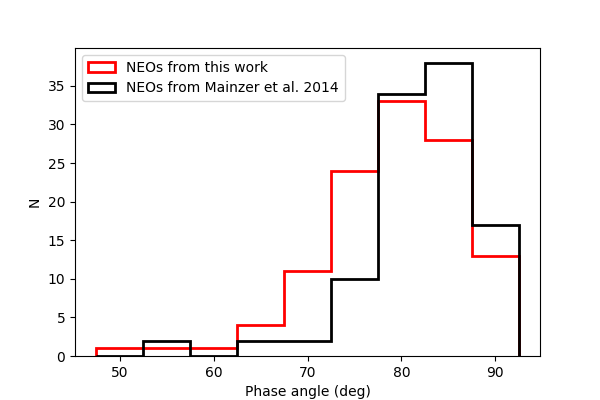}
\protect\caption{Distribution of observed phase angles for all NEOs
  presented here (red) and those from \citet{mainzer14tinyneo}
  (black).}
\label{fig.phase}
\end{center}
\end{figure}

Using our W1 and W2 measurements, inferred $V$ magnitude, heliocentric
and spacecraft-centric distances, phase angle, and assumed infrared
albedo ratio ($\frac{p_{IR}}{p_V}=1.6 \pm 1.0$) we simultaneously
constrain the diameter and V-band albedo for each object using a
least-squared minimizer available through the {\it scipy} Python
package\footnote{\it https://www.scipy.org} \citep{scipy}.  We set a
limit of $\sigma<0.25~$mag on the uncertainty of the measured
magnitude in each band for it to be used in the thermal fitting;
detections with larger measurement uncertainties were not used as
model constraints.  If the W1 and W2 fluxes are both dominated by
thermal emission (i.e. the reflected light component is $<10\%$) we
can constrain the beaming parameter.  For objects detected in only a
single band, or that are dominated by reflected light in W1, we assume
a beaming parameter ($\eta$) of $2.0$.  The assumed beaming parameter
we choose here is larger than what was assumed in other thermal
modeling papers of NEOWISE data
\citep[e.g.][]{nugent15,nugent16,masiero17}, and is instead drawn from
the mean of the fitted beaming parameters for small NEOs found in
\citet{mainzer14tinyneo}.  We show in Figure~\ref{fig.phase} that the
distribution of phase angles for our objects is comparable to those
from \citet{mainzer14tinyneo}, and thus the larger assumed beaming
parameter is appropriate because these close-pass objects are at
higher phases on average and in general smaller than NEOs detected by
WMOPS \citet{wolters09,mainzer11neo}.

In order to determine uncertainties on our fitted physical properties,
we perform 25 Monte Carlo simulations of each fit, varying the
measured magnitudes within their measurement uncertainties, and
assuming a random error of $0.2$ magnitudes for $H$, $0.5$ for $\eta$
(when not fit), and $1.0$ for the infrared albedo ratio.  These
parameters match our standard assumptions for NEOs from previous work
\citep[cf.][]{mainzer11neo,mainzer14tinyneo}, and are used to
encapsulate the range of values these parameters are observed to have.
In addition to the random error determined through Monte Carlo trials,
we add an additional diameter uncertainty due to the incomplete
rotational phase coverage for objects with a small number of
detections (see below).

We reject from our fits any object where the final reflected light
component in W2 was $>10\%$ of the total flux.  Thermal fits to W2 can
be significantly altered by assumptions of the NIR albedo, and lead
to unstable solutions.  We have found in previous work that a cut at
$10\%$ reflected light results in a more reliable list of physical
properties \citep{masiero17}.  Table~\ref{tab.fits} gives our fitted
diameters and albedos for the 116 near-Earth objects.

\subsection{Diameter Uncertainty}

Thermal modeling using NEATM results in a constraint on the equivalent
circular diameter of the projection onto the sky of the true 3D shape
of the observed asteroid at the time of observation.  If the object
has been observed at multiple different rotation phases, the best
simultaneous fit of all observations thus corresponds to the
spherical-equivalent flux for that viewing geometry.  Multiple
observing epochs can reduce the uncertainty due to pole-on vs
equator-on viewing geometries.  However, for objects observed only a
few times, or even a single time, the fit would instead constrain a
sphere with proportions set by the effective circular area of the
object at the time of observation.  Thus, the uncertainties on
diameter for the fits presented here tend to be larger than for
objects with better coverage, even when both bands are thermally
dominated and a beaming parameter can be fit.  For nearly circular
objects this additional error will be small, while for very elongated
shapes it can be the dominant source of error.

\begin{figure}[ht]
\begin{center}
\includegraphics[scale=0.5]{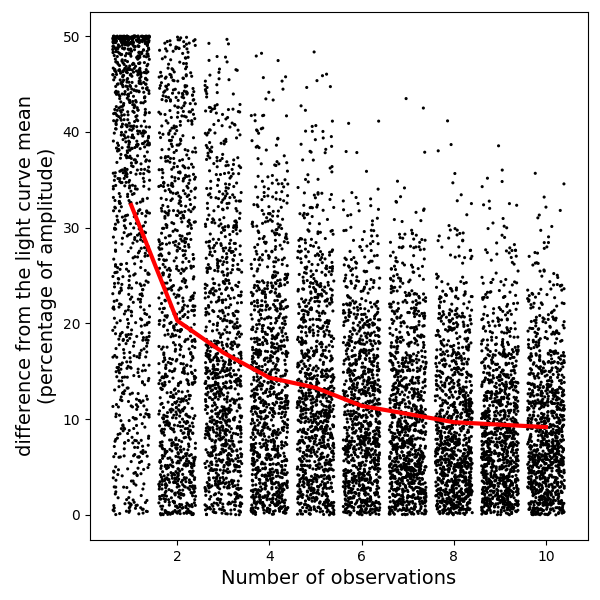}
\protect\caption{Outcome of a Monte Carlo simulation of $N$ randomly
  spaced observations of a theoretical asteroid light curve, showing
  the resulting deviation between the measured mean value and the
  actual light curve mean as a percentage of amplitude $A$.  Each
  point is given an offset in the range of (-0.4,0.4) from the integer
  $N$ observations for clarity.  The average of all trials (red line)
  decreases as the square-root of number of observations of the light
  curve.  }
\label{fig.mclc}
\end{center}
\end{figure}

The error due to unknown rotation phase is a function of the number of
detections, and the amplitude of the light curve. To test the effect
of the unknown light curve on the measured flux, we perform a Monte
Carlo simulation of $N$ random samples of a theoretical light curve
with amplitude $A$.We compare the average from those $N$ samples to
the true mean of the light curve, and characterize the light
curve-induced deviation as a fraction of $A$.  As light curve
amplitude is measured peak-to-trough, the maximum possible deviation
is $50\%~A$, while on average a single observation will result in a
mean deviation of $32\%~A$ from the midpoint of the light curve.  As
the number of observations $N$ increases, and the light curve is
better sampled, the mean deviation decreases.  We show in
Figure~\ref{fig.mclc} the results of our Monte Carlo simulation for a
range of $N$ values.  These simulations make the simplifying
assumption that the light curve follows a basic sinusoidal profile as
would be expected for a rotating extended triaxial ellipsoid.  More
realistic light curve shapes with asymmetries could potentially alter
the results if these deviations changed the fraction of the light
curve that was near the extrema. However, the unknown light curve
amplitude is a much more dominant source of uncertainty in this
analysis than the non-sinusoidal shape or the range of amplitude
deviations seen in Figure~\ref{fig.mclc}.

In order to translate this deviation from a percentage of amplitude
$A$ to a measured magnitude deviation, we use the observed light curve
amplitudes of all NEOs automatically detected by NEOWISE during the
first three years of the Reactivation survey
\citep{nugent15,nugent16,masiero17}.  The mean observed W2 magnitude
variation (an analog for light curve amplitude) is $0.57~$mag, and the
median is $0.5~$mag.  These numbers are the same if we consider only
the 32 NEOs smaller than $D<250~m$.  Therefore a single observation
has a typical uncertainty of $32\%~A = 0.32 \times
0.57~$mag$\sim0.2~$mag from the light curve mean, which corresponds to
$\sim20\%$ in flux or $\sim10\%$ in diameter.  Using the maximum
observed magnitude variation of $A\sim1.8~mag$ would result in an
uncertainty of $\sim60\%$ in flux or $\sim30\%$ in
diameter. Conversely, an object observed pole-on, or that is
spherical, would present no apparent light curve variations and thus
no additional uncertainty from the small number of observations.
However without more data on the spin state of each object, we cannot
place a better constraint on this uncertainty, and thus default to the
mean value.

This uncertainty decreases with the square-root of the number of
observations as shown by the average of all simulations in
Figure~\ref{fig.mclc}.  As noted in \citet{mainzer14tinyneo}, the true
value of this error for an individual object is highly dependent on
the actual light curve amplitude, triaxial shape, and viewing
geometry.  Additionally, this error is not distributed normally around
the best-fit diameter, as an object is more likely to be seen at the
brightest excursion from its true mean flux as noted by the analysis
from the Infrared Astronomical Satellite
mission\footnote{https://irsa.ipac.caltech.edu/IRASdocs/exp.sup/ch11/J.html}.
Thus, the diameters presented here will preferentially over-estimate
the true size of these NEOs.  We include the typical additional
uncertainty in quadrature with the diameter error calculated from the
Monte Carlo trials, but note that there could be a much larger
additional diameter error of order $30\%$ for specific cases.  We
estimate this component of the error ($\sigma_{LC}$) based on the
number of observations in W2 ($n_{obs}$), using the equation:

\[\sigma_{LC} = \frac{0.10 D}{\sqrt{n_{obs}}}\]

and we propagate this error to the error on albedo as well.

Recently, \citet{mommert18} have shown that the NEATM model does not perform
as well at phase angles larger than $\alpha>65^\circ$ as it does for
smaller phase angles, and frequently overestimates the diameter of
NEOs seen at these phases.  This will be an additional source of
uncertainty on top of the incomplete light curve sampling.  As nearly
all of the objects presented in this paper are at large phase angles,
we apply the high-phase NEATM diameter and albedo correction terms
from \citet{mommert18} to our fits, and present these as separate
columns in Table~\ref{tab.fits}.

\section{Results and Discussion}

Of the $122$ objects that were detected, $5$ thermal fits were
rejected due to a reflected light contribution of $>10\%$ in the W2
band.  One object, 2016 TB57, was not fit because it has only two
detections, both of which had magnitude errors larger than the
threshold of $0.25~$mag, and thus insufficient for constraining the
thermal emission.

\begin{figure}[ht]
\begin{center}
\includegraphics[scale=0.5]{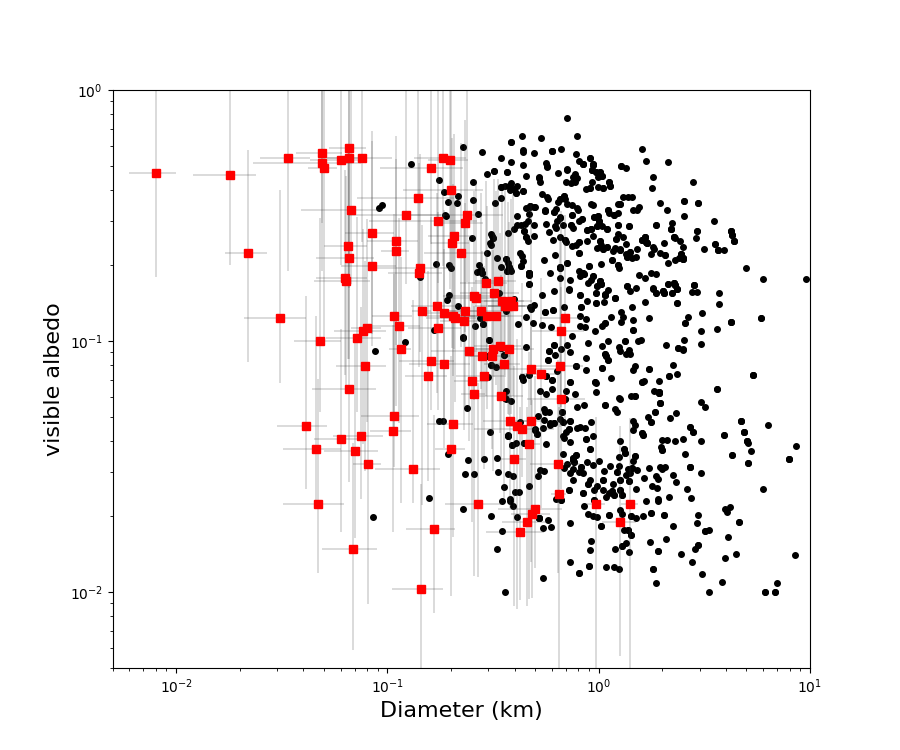}
\protect\caption{Comparison of fitted diameters and albedos for the
  short-arc objects presented here (red squares) with the NEOs found
  by the automated detection routines during the first three years of
  the reactivated NEOWISE survey (black).  The short-arc objects tend
  to be smaller than objects detected automatically, and show a
  significant bias against small, low albedo asteroids that is a result
  of the visible-light selection effects imposed by the
  ground-based surveys that discovered this population.  Error bars on
  previously reported objects are omitted for clarity. }
\label{fig.diamalb}
\end{center}
\end{figure}

Thermal infrared fits for $116$ small NEOs are presented in Table
\ref{tab.fits}.  We compare the diameters and albedos found here with
those of all NEOs measured that were detected by WMOPS during the
reactivated NEOWISE mission \citep{nugent15,nugent16,masiero17} in
Figure~\ref{fig.diamalb}.  The short-arc objects we report here tend
to be significantly smaller than the objects detected by WMOPS, as
would be expected for objects passing close to the Earth and thus
moving quickly across the field of view.  Additionally, we see
imprinted in the population a bias against small, low albedo NEOs.
This is an artifact of the flux-limited Malmquist bias that impacts
the ground-based visible light surveys that discovered the short-arc
objects.  In contrast, the NEOWISE-detected objects show a nearly
uniform sensitivity with respect to albedo, as observed for previous
phases of the WISE survey \citep{mainzer11neo}.

\begin{figure}[ht]
\begin{center}
\includegraphics[scale=0.5]{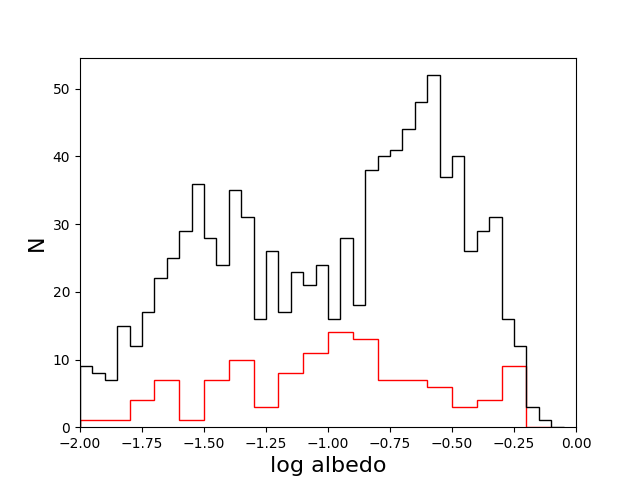}
\protect\caption{Histogram of the albedos of the short-arc objects
  reported here (red) and previously reported NEOs characterized by
  NEOWISE (black).  Note the bins for newly reported objects are twice
  as wide as those for previously reported NEOs.  
 }
\label{fig.albhist}
\end{center}
\end{figure}

Looking at the albedos alone, we show in Figure~\ref{fig.albhist} the
distributions of albedos for both the short-arc population reported
here and the population of objects previously reported.  While the
histogram of previously-reported objects shows the expected bimodal
albedo distribution \citep{mainzer11neo}, the short-arc objects are
skewed toward higher albedos when compared with the population
selected based on W2 flux.  Although the distribution of
  albedos in this work have relatively few objects in each bin, a KS
  test allows us to reject the assumption that they are drawn from the
  same population at the $>98\%$ level.  The difference albedo
  distributions is likely an artifact of the incomplete rotation
phase coverage coupled with the Eddington bias favoring detection at
an object's brighter apparition.  For these cases, the NEOWISE
diameter fit will be systematically greater than the true effective
spherical diameter, which will result in a shift in fitted albedo to
lower values (for a fixed absolute magnitude).  Thus, the population
of objects with $p_V\sim10\%$ likely will have over-estimated
diameters and under-estimated albedos, which would shift them to the
bright peak in the albedo histogram.  This shift would be in addition
to the phase-correction to NEATM.  However, as we cannot know which
specific objects are suffering from this bias without other data, we
cannot correct for it.  Instead, we encourage caution in
interpretation of these results, especially when comparing to results
drawn from the WMOPS-detected sample, which is selected based on the
thermal emission-dominated W2 flux as opposed to reflected visible
light flux.

\clearpage

\section{Conclusions}

We present diameter fits for $116$ short-arc NEOs seen during the
first three years of the reactivated NEOWISE survey.  Combined with
the $541$ unique near-Earth objects that were automatically detected
and characterized over that time, NEOWISE has provided
characterization data for $657$ NEOs since the survey was restarted in
2013.  This bring the total count of NEOs with thermal
infrared characterization from all phases of the WISE and NEOWISE
surveys to $1203$, which represents $7\%$ of the known NEO population
at the time of writing.  The analysis presented here shows the utility
of recording and archiving full-frame images from surveys, as it
allows later searches for object that were missed during automatic
processing, greatly enhancing the scientific return of these data.

\section*{Acknowledgments}

We thank the referee for their detailed and helpful comments that
improved the manuscript.  This research was carried out at the Jet
Propulsion Laboratory, California Institute of Technology, under a
contract with the National Aeronautics and Space Administration.  This
publication makes use of data products from the Wide-field Infrared
Survey Explorer, which is a joint project of the University of
California, Los Angeles, and the Jet Propulsion Laboratory/California
Institute of Technology, funded by the National Aeronautics and Space
Administration.  This publication also makes use of data products from
NEOWISE, which is a project of the Jet Propulsion
Laboratory/California Institute of Technology, funded by the Planetary
Science Division of the National Aeronautics and Space Administration.
This research has made use of data and services provided by the
International Astronomical Union's Minor Planet Center.  This research
has made use of the NASA/IPAC Infrared Science Archive, which is
operated by the California Institute of Technology, under contract
with the National Aeronautics and Space Administration.

\vspace{0.5in}

\scriptsize{
\begin{longtable}{cccccccccccc}
\caption{Thermal model fits for short arc NEOs observed in the first
  three years of the NEOWISE Reactivation survey. Names are in
  MPC-packed format, H and G are the photometric parameters, $p_V$ is
  the visible light albedo, n$_{W1}$ and n$_{W2}$ are the numbers of
  detections in the W1 and W2 bandpasses, the Fitted Beaming column is
  '1' if the beaming parameter was fitted during the NEATM modeling
  and '0' if an assumed value was used. $D_{corr}$ and $pV_{corr}$ are
  the phase-corrected diameter using the equations from
  \citet{mommert18}. }\\

  Name  &  H  &   G   &   Diameter  &  D$_{corr}$ & $\log{p_V}$  & $\log{p_V}_{corr}$ &  beaming  &  n$_{W1}$  &n$_{W2}$ & phase & Fitted Beaming? \\
    &  (mag)  &      &   (km)  &  (km)  &  & &   &   & &  (deg) &  \\
  \hline
  \endhead
K14A00C & 21.00 &  0.15 & 0.343 $\pm$ 0.094 & 0.316 & -1.22 $\pm$ 0.31 & -1.15 & 2.00 $\pm$ 1.00 &   0 &   1 & 71.37 & 0\\
K14A16G & 22.70 &  0.15 & 0.200 $\pm$ 0.032 & 0.165 & -1.43 $\pm$ 0.13 & -1.28 & 1.76 $\pm$ 0.23 &   4 &   4 & 86.35 & 1\\
K14A51N & 20.50 &  0.15 & 0.222 $\pm$ 0.073 & 0.193 & -0.65 $\pm$ 0.24 & -0.53 & 2.00 $\pm$ 1.00 &   0 &   1 & 80.29 & 0\\
K14B02X & 21.90 &  0.15 & 0.110 $\pm$ 0.030 & 0.097 & -0.60 $\pm$ 0.21 & -0.49 & 0.67 $\pm$ 0.20 &   2 &   2 & 78.46 & 1\\
K14B08R & 21.70 &  0.15 & 0.141 $\pm$ 0.040 & 0.118 & -0.73 $\pm$ 0.24 & -0.59 & 2.00 $\pm$ 1.00 &   0 &   3 & 84.57 & 0\\
K14B25H & 21.60 &  0.15 & 0.257 $\pm$ 0.033 & 0.219 & -1.21 $\pm$ 0.09 & -1.08 & 1.25 $\pm$ 0.13 &   3 &   3 & 82.58 & 1\\
K14C00R & 22.50 &  0.15 & 0.116 $\pm$ 0.014 & 0.094 & -1.03 $\pm$ 0.12 & -0.86 & 2.95 $\pm$ 0.15 &   1 &   1 & 88.02 & 1\\
K14C13D & 21.30 &  0.15 & 0.398 $\pm$ 0.055 & 0.338 & -1.47 $\pm$ 0.13 & -1.34 & 1.30 $\pm$ 0.10 &   1 &   1 & 82.85 & 1\\
K14F00D & 25.50 &  0.15 & 0.022 $\pm$ 0.005 & 0.018 & -0.65 $\pm$ 0.20 & -0.48 & 2.00 $\pm$ 1.00 &   0 &   1 & 89.08 & 0\\
K14G35J & 20.30 &  0.15 & 0.377 $\pm$ 0.115 & 0.351 & -1.03 $\pm$ 0.34 & -0.96 & 2.00 $\pm$ 1.00 &   0 &   2 & 69.88 & 0\\
K14H04G & 19.80 &  0.15 & 0.975 $\pm$ 0.099 & 0.883 & -1.65 $\pm$ 0.09 & -1.56 & 1.76 $\pm$ 0.13 &   4 &   4 & 74.56 & 1\\
K14J55V & 20.10 &  0.15 & 0.321 $\pm$ 0.100 & 0.290 & -0.81 $\pm$ 0.27 & -0.72 & 2.00 $\pm$ 1.00 &   0 &   2 & 74.82 & 0\\
K14K76O & 23.90 &  0.15 & 0.166 $\pm$ 0.043 & 0.141 & -1.75 $\pm$ 0.27 & -1.62 & 2.00 $\pm$ 1.00 &   0 &   4 & 82.73 & 0\\
K14K86V & 20.30 &  0.15 & 0.644 $\pm$ 0.173 & 0.618 & -1.49 $\pm$ 0.20 & -1.45 & 2.00 $\pm$ 1.00 &   0 &   2 & 63.68 & 0\\
K14M05S & 22.70 &  0.15 & 0.113 $\pm$ 0.030 & 0.093 & -0.94 $\pm$ 0.23 & -0.79 & 2.00 $\pm$ 1.00 &   0 &   1 & 85.82 & 0\\
K14M18F & 26.00 &  0.15 & 0.069 $\pm$ 0.020 & 0.057 & -1.83 $\pm$ 0.22 & -1.68 & 2.00 $\pm$ 1.00 &   0 &   5 & 85.18 & 0\\
K14N03E & 20.10 &  0.15 & 0.233 $\pm$ 0.051 & 0.202 & -0.53 $\pm$ 0.20 & -0.41 & 2.00 $\pm$ 1.00 &   0 &   2 & 80.82 & 0\\
K14N52K & 21.30 &  0.15 & 0.502 $\pm$ 0.170 & 0.462 & -1.67 $\pm$ 0.38 & -1.60 & 2.00 $\pm$ 1.00 &   0 &   4 & 71.77 & 0\\
K14P59S & 20.50 &  0.15 & 0.206 $\pm$ 0.051 & 0.181 & -0.58 $\pm$ 0.19 & -0.47 & 2.00 $\pm$ 1.00 &   0 &   5 & 78.84 & 0\\
K14P59W & 20.90 &  0.15 & 0.412 $\pm$ 0.048 & 0.352 & -1.34 $\pm$ 0.09 & -1.21 & 1.41 $\pm$ 0.10 &   2 &   2 & 82.49 & 1\\
K14QT5T & 25.90 &  0.15 & 0.041 $\pm$ 0.011 & 0.033 & -1.34 $\pm$ 0.36 & -1.18 & 2.00 $\pm$ 1.00 &   0 &   2 & 87.56 & 0\\
K14SE1V & 21.60 &  0.15 & 0.172 $\pm$ 0.043 & 0.149 & -0.86 $\pm$ 0.26 & -0.75 & 2.00 $\pm$ 1.00 &   0 &   8 & 80.37 & 0\\
K14SQ0U & 21.80 &  0.15 & 0.174 $\pm$ 0.049 & 0.149 & -0.95 $\pm$ 0.28 & -0.83 & 2.00 $\pm$ 1.00 &   0 &   2 & 81.36 & 0\\
K14V02G & 22.70 &  0.15 & 0.085 $\pm$ 0.025 & 0.073 & -0.70 $\pm$ 0.33 & -0.57 & 2.00 $\pm$ 1.00 &   0 &   5 & 82.87 & 0\\
K14V02K & 18.70 &  0.15 & 0.691 $\pm$ 0.161 & 0.648 & -0.91 $\pm$ 0.20 & -0.85 & 2.00 $\pm$ 1.00 &   0 &   6 & 68.40 & 0\\
K14V06L & 21.30 &  0.15 & 0.242 $\pm$ 0.081 & 0.211 & -1.04 $\pm$ 0.25 & -0.93 & 1.54 $\pm$ 0.43 &   3 &   3 & 80.16 & 1\\
K14W06L & 20.00 &  0.15 & 0.198 $\pm$ 0.043 & 0.176 & -0.28 $\pm$ 0.16 & -0.18 & 1.24 $\pm$ 0.29 &   3 &   2 & 77.51 & 1\\
K14WC0Z & 20.50 &  0.15 & 0.297 $\pm$ 0.078 & 0.245 & -0.90 $\pm$ 0.24 & -0.75 & 2.06 $\pm$ 0.47 &   2 &   2 & 85.93 & 1\\
K14Wa8K & 21.40 &  0.15 & 0.486 $\pm$ 0.103 & 0.441 & -1.69 $\pm$ 0.27 & -1.60 & 2.00 $\pm$ 1.00 &   0 &   3 & 74.14 & 0\\
K14Y14L & 20.60 &  0.15 & 0.355 $\pm$ 0.067 & 0.304 & -1.09 $\pm$ 0.15 & -0.97 & 1.68 $\pm$ 0.24 &   1 &   1 & 81.78 & 1\\
K15BV0W & 22.00 &  0.15 & 0.146 $\pm$ 0.015 & 0.126 & -0.88 $\pm$ 0.08 & -0.76 & 0.89 $\pm$ 0.08 &   4 &   4 & 81.16 & 1\\
K15Bo9N & 20.70 &  0.15 & 0.315 $\pm$ 0.067 & 0.274 & -1.03 $\pm$ 0.17 & -0.91 & 2.00 $\pm$ 1.00 &   0 &   4 & 80.17 & 0\\
K15C12X & 20.00 &  0.15 & 0.184 $\pm$ 0.051 & 0.154 & -0.27 $\pm$ 0.20 & -0.13 & 0.85 $\pm$ 0.25 &   1 &   1 & 84.54 & 1\\
K15C13P & 19.80 &  0.15 & 0.384 $\pm$ 0.098 & 0.334 & -0.84 $\pm$ 0.19 & -0.73 & 1.54 $\pm$ 0.34 &   4 &   4 & 80.25 & 1\\
K15D54A & 20.80 &  0.15 & 0.468 $\pm$ 0.058 & 0.392 & -1.41 $\pm$ 0.12 & -1.27 & 1.18 $\pm$ 0.07 &   1 &   1 & 84.57 & 1\\
K15DI0U & 20.80 &  0.15 & 0.435 $\pm$ 0.179 & 0.397 & -1.35 $\pm$ 0.45 & -1.27 & 2.00 $\pm$ 1.00 &   0 &   1 & 73.33 & 0\\
K15DL5N & 19.80 &  0.15 & 0.535 $\pm$ 0.109 & 0.453 & -1.13 $\pm$ 0.15 & -1.00 & 1.88 $\pm$ 0.30 &   2 &   2 & 83.46 & 1\\
K15E07D & 20.80 &  0.15 & 0.313 $\pm$ 0.068 & 0.278 & -1.06 $\pm$ 0.24 & -0.96 & 2.00 $\pm$ 1.00 &   0 &   1 & 77.15 & 0\\
K15E07E & 20.20 &  0.15 & 0.294 $\pm$ 0.074 & 0.262 & -0.77 $\pm$ 0.19 & -0.67 & 2.00 $\pm$ 1.00 &   0 &   6 & 77.05 & 0\\
K15F00L & 20.80 &  0.15 & 0.232 $\pm$ 0.040 & 0.197 & -0.88 $\pm$ 0.23 & -0.75 & 2.75 $\pm$ 0.38 &   2 &   2 & 82.70 & 1\\
K15F33S & 22.20 &  0.15 & 0.066 $\pm$ 0.014 & 0.054 & -0.27 $\pm$ 0.16 & -0.11 & 0.90 $\pm$ 0.24 &   1 &   2 & 87.52 & 1\\
K15FC0N & 23.50 &  0.15 & 0.064 $\pm$ 0.019 & 0.056 & -0.76 $\pm$ 0.21 & -0.65 & 0.96 $\pm$ 0.27 &   1 &   1 & 80.06 & 1\\
K15G00S & 20.60 &  0.15 & 0.262 $\pm$ 0.063 & 0.232 & -0.83 $\pm$ 0.29 & -0.73 & 2.00 $\pm$ 1.00 &   0 &   2 & 77.53 & 0\\
K15H01F & 22.70 &  0.15 & 0.108 $\pm$ 0.023 & 0.095 & -0.90 $\pm$ 0.19 & -0.79 & 2.00 $\pm$ 1.00 &   0 &   3 & 78.78 & 0\\
K15HH1U & 23.70 &  0.15 & 0.108 $\pm$ 0.033 & 0.089 & -1.30 $\pm$ 0.43 & -1.14 & 2.00 $\pm$ 1.00 &   0 &   4 & 86.91 & 0\\
K15HI1X & 23.90 &  0.15 & 0.078 $\pm$ 0.021 & 0.064 & -1.10 $\pm$ 0.21 & -0.94 & 2.00 $\pm$ 1.00 &   0 &   1 & 87.52 & 0\\
K15J02C & 21.00 &  0.15 & 0.382 $\pm$ 0.099 & 0.350 & -1.32 $\pm$ 0.26 & -1.24 & 2.00 $\pm$ 1.00 &   0 &   4 & 72.71 & 0\\
K15KC2N & 20.50 &  0.15 & 0.342 $\pm$ 0.086 & 0.311 & -1.02 $\pm$ 0.26 & -0.94 & 2.00 $\pm$ 1.00 &   0 &   4 & 73.84 & 0\\
K15L21L & 19.90 &  0.15 & 0.374 $\pm$ 0.074 & 0.331 & -0.86 $\pm$ 0.20 & -0.76 & 1.19 $\pm$ 0.23 &   3 &   3 & 77.91 & 1\\
K15L21M & 22.10 &  0.15 & 0.066 $\pm$ 0.013 & 0.056 & -0.23 $\pm$ 0.14 & -0.10 & 0.71 $\pm$ 0.15 &   1 &   1 & 82.20 & 1\\
K15MB6N & 19.80 &  0.15 & 0.394 $\pm$ 0.095 & 0.365 & -0.86 $\pm$ 0.25 & -0.79 & 2.00 $\pm$ 1.00 &   0 &   8 & 71.06 & 0\\
K15N13Z & 20.60 &  0.15 & 0.258 $\pm$ 0.075 & 0.219 & -0.82 $\pm$ 0.21 & -0.68 & 1.48 $\pm$ 0.38 &   1 &   1 & 83.06 & 1\\
K15P00C & 19.40 &  0.15 & 1.266 $\pm$ 0.230 & 1.159 & -1.72 $\pm$ 0.15 & -1.64 & 2.21 $\pm$ 0.30 &   4 &   4 & 72.89 & 1\\
K15P57K & 24.70 &  0.15 & 0.075 $\pm$ 0.020 & 0.062 & -1.38 $\pm$ 0.36 & -1.23 & 2.00 $\pm$ 1.00 &   0 &   3 & 86.34 & 0\\
K15PM8U & 20.30 &  0.15 & 0.328 $\pm$ 0.106 & 0.282 & -0.90 $\pm$ 0.33 & -0.78 & 2.00 $\pm$ 1.00 &   0 &   1 & 81.58 & 0\\
K15Q00G & 23.80 &  0.15 & 0.132 $\pm$ 0.045 & 0.109 & -1.51 $\pm$ 0.57 & -1.36 & 2.00 $\pm$ 1.00 &   0 &   1 & 85.55 & 0\\
K15T00E & 22.50 &  0.15 & 0.156 $\pm$ 0.035 & 0.128 & -1.14 $\pm$ 0.21 & -0.98 & 1.46 $\pm$ 0.26 &   1 &   1 & 86.66 & 1\\
K15T24X & 21.50 &  0.15 & 0.252 $\pm$ 0.046 & 0.208 & -1.16 $\pm$ 0.14 & -1.00 & 1.27 $\pm$ 0.19 &   3 &   3 & 86.05 & 1\\
K15TN8K & 21.90 &  0.15 & 0.076 $\pm$ 0.029 & 0.067 & -0.27 $\pm$ 0.27 & -0.16 & 0.90 $\pm$ 0.35 &   1 &   1 & 78.54 & 1\\
K15TW3D & 19.90 &  0.15 & 0.332 $\pm$ 0.072 & 0.307 & -0.76 $\pm$ 0.19 & -0.68 & 2.00 $\pm$ 1.00 &   0 &   2 & 71.40 & 0\\
K15U67M & 18.90 &  0.15 & 0.665 $\pm$ 0.183 & 0.577 & -0.96 $\pm$ 0.30 & -0.84 & 2.35 $\pm$ 0.57 &   2 &   2 & 80.36 & 1\\
K15V01E & 21.00 &  0.15 & 0.139 $\pm$ 0.067 & 0.112 & -0.43 $\pm$ 0.34 & -0.26 & 1.35 $\pm$ 0.65 &   3 &   3 & 88.50 & 1\\
K15V01F & 23.80 &  0.15 & 0.072 $\pm$ 0.021 & 0.059 & -0.99 $\pm$ 0.31 & -0.83 & 2.00 $\pm$ 1.00 &   0 &   2 & 86.48 & 0\\
K15V65B & 22.90 &  0.15 & 0.049 $\pm$ 0.026 & 0.042 & -0.29 $\pm$ 0.37 & -0.16 & 0.91 $\pm$ 0.53 &   2 &   2 & 82.11 & 1\\
K15V65U & 26.40 &  0.15 & 0.047 $\pm$ 0.015 & 0.037 & -1.65 $\pm$ 0.33 & -1.48 & 2.00 $\pm$ 1.00 &   0 &   1 & 89.35 & 0\\
K15W09G & 20.30 &  0.15 & 0.480 $\pm$ 0.062 & 0.412 & -1.32 $\pm$ 0.14 & -1.20 & 3.14 $\pm$ 0.23 &   1 &   1 & 81.88 & 1\\
K15W13H & 19.00 &  0.15 & 1.403 $\pm$ 0.129 & 1.305 & -1.65 $\pm$ 0.08 & -1.58 & 1.65 $\pm$ 0.10 &   3 &   3 & 70.20 & 1\\
K15X00E & 24.70 &  0.15 & 0.048 $\pm$ 0.012 & 0.042 & -1.00 $\pm$ 0.32 & -0.89 & 2.00 $\pm$ 1.00 &   0 &   4 & 79.37 & 0\\
K15X01D & 20.10 &  0.15 & 0.201 $\pm$ 0.082 & 0.177 & -0.40 $\pm$ 0.29 & -0.29 & 0.95 $\pm$ 0.41 &   3 &   3 & 78.42 & 1\\
K15X01K & 20.00 &  0.15 & 0.350 $\pm$ 0.092 & 0.325 & -0.84 $\pm$ 0.24 & -0.77 & 2.00 $\pm$ 1.00 &   0 &   6 & 70.50 & 0\\
K15XC9O & 25.20 &  0.15 & 0.060 $\pm$ 0.015 & 0.053 & -1.39 $\pm$ 0.24 & -1.29 & 2.00 $\pm$ 1.00 &   0 &   2 & 78.85 & 0\\
K15Y00K & 25.90 &  0.15 & 0.046 $\pm$ 0.014 & 0.038 & -1.43 $\pm$ 0.38 & -1.28 & 2.00 $\pm$ 1.00 &   0 &   5 & 86.51 & 0\\
K15Y01B & 21.50 &  0.15 & 0.186 $\pm$ 0.047 & 0.161 & -0.89 $\pm$ 0.20 & -0.77 & 2.00 $\pm$ 1.00 &   0 &   2 & 80.73 & 0\\
K15Y07X & 22.00 &  0.15 & 0.110 $\pm$ 0.017 & 0.095 & -0.64 $\pm$ 0.12 & -0.52 & 1.09 $\pm$ 0.17 &   6 &   7 & 81.46 & 1\\
K15Y10T & 20.00 &  0.15 & 0.359 $\pm$ 0.099 & 0.339 & -0.86 $\pm$ 0.32 & -0.81 & 2.00 $\pm$ 1.00 &   0 &   7 & 67.38 & 0\\
K16AG6D & 23.60 &  0.15 & 0.077 $\pm$ 0.021 & 0.063 & -0.96 $\pm$ 0.19 & -0.80 & 2.00 $\pm$ 1.00 &   0 &   1 & 87.23 & 0\\
K16B14P & 21.20 &  0.15 & 0.285 $\pm$ 0.073 & 0.264 & -1.14 $\pm$ 0.24 & -1.07 & 2.00 $\pm$ 1.00 &   0 &   5 & 70.57 & 0\\
K16B15J & 23.30 &  0.15 & 0.080 $\pm$ 0.013 & 0.064 & -0.95 $\pm$ 0.24 & -0.77 & 2.76 $\pm$ 0.41 &   2 &   2 & 89.01 & 1\\
K16B80R & 21.30 &  0.15 & 0.208 $\pm$ 0.056 & 0.181 & -0.91 $\pm$ 0.27 & -0.80 & 2.00 $\pm$ 1.00 &   0 &   4 & 79.97 & 0\\
K16C30U & 21.60 &  0.15 & 0.459 $\pm$ 0.111 & 0.435 & -1.72 $\pm$ 0.27 & -1.66 & 2.00 $\pm$ 1.00 &   0 &   5 & 66.42 & 0\\
K16C31B & 25.00 &  0.15 & 0.070 $\pm$ 0.020 & 0.056 & -1.44 $\pm$ 0.26 & -1.26 & 2.00 $\pm$ 1.00 &   0 &   2 & 89.10 & 0\\
K16CD6A & 19.30 &  0.15 & 0.654 $\pm$ 0.193 & 0.649 & -1.10 $\pm$ 0.37 & -1.09 & 2.00 $\pm$ 1.00 &   0 &  13 & 52.31 & 0\\
K16CD6L & 21.40 &  0.15 & 0.123 $\pm$ 0.057 & 0.113 & -0.50 $\pm$ 0.33 & -0.42 & 1.50 $\pm$ 0.67 &   3 &   3 & 72.90 & 1\\
K16D02O & 23.70 &  0.15 & 0.034 $\pm$ 0.009 & 0.031 & -0.27 $\pm$ 0.19 & -0.18 & 0.93 $\pm$ 0.30 &   2 &   2 & 75.26 & 1\\
K16E01E & 24.50 &  0.15 & 0.066 $\pm$ 0.022 & 0.055 & -1.19 $\pm$ 0.24 & -1.05 & 2.10 $\pm$ 0.54 &   1 &   1 & 84.89 & 1\\
K16E01V & 23.90 &  0.15 & 0.106 $\pm$ 0.023 & 0.095 & -1.36 $\pm$ 0.22 & -1.27 & 2.00 $\pm$ 1.00 &   0 &   2 & 76.55 & 0\\
K16E26Z & 22.60 &  0.15 & 0.268 $\pm$ 0.080 & 0.227 & -1.65 $\pm$ 0.31 & -1.51 & 2.00 $\pm$ 1.00 &   0 &   6 & 83.45 & 0\\
K16EF7H & 21.10 &  0.15 & 0.231 $\pm$ 0.054 & 0.213 & -0.92 $\pm$ 0.20 & -0.84 & 2.00 $\pm$ 1.00 &   0 &   5 & 71.61 & 0\\
K16F03Y & 21.30 &  0.15 & 0.205 $\pm$ 0.055 & 0.163 & -0.90 $\pm$ 0.37 & -0.72 & 2.49 $\pm$ 0.59 &   1 &   1 & 90.16 & 1\\
K16F12E & 20.60 &  0.15 & 0.202 $\pm$ 0.042 & 0.171 & -0.61 $\pm$ 0.21 & -0.47 & 2.00 $\pm$ 1.00 &   0 &   3 & 83.84 & 0\\
K16F13C & 22.00 &  0.15 & 0.186 $\pm$ 0.060 & 0.163 & -1.09 $\pm$ 0.27 & -0.98 & 2.00 $\pm$ 1.00 &   0 &   4 & 78.72 & 0\\
K16GM0P & 21.90 &  0.15 & 0.422 $\pm$ 0.128 & 0.406 & -1.76 $\pm$ 0.33 & -1.72 & 2.00 $\pm$ 1.00 &   0 &   2 & 63.10 & 0\\
K16J17M & 19.60 &  0.15 & 0.661 $\pm$ 0.205 & 0.648 & -1.23 $\pm$ 0.25 & -1.21 & 2.00 $\pm$ 1.00 &   0 &   3 & 57.25 & 0\\
K16J28W & 25.40 &  0.15 & 0.031 $\pm$ 0.010 & 0.029 & -0.91 $\pm$ 0.35 & -0.83 & 2.00 $\pm$ 1.00 &   0 &   1 & 73.71 & 0\\
K16J33U & 20.70 &  0.15 & 0.174 $\pm$ 0.092 & 0.156 & -0.52 $\pm$ 0.36 & -0.42 & 1.32 $\pm$ 0.65 &   2 &   2 & 75.98 & 1\\
K16K00D & 22.30 &  0.15 & 0.160 $\pm$ 0.046 & 0.144 & -1.08 $\pm$ 0.44 & -0.99 & 2.00 $\pm$ 1.00 &   0 &   4 & 75.78 & 0\\
K16L00B & 22.80 &  0.15 & 0.060 $\pm$ 0.017 & 0.048 & -0.28 $\pm$ 0.21 & -0.11 & 1.28 $\pm$ 0.40 &   4 &   4 & 87.95 & 1\\
K16L02F & 22.40 &  0.15 & 0.204 $\pm$ 0.050 & 0.189 & -1.33 $\pm$ 0.19 & -1.26 & 2.00 $\pm$ 1.00 &   0 &   9 & 71.09 & 0\\
K16L09D & 22.40 &  0.15 & 0.085 $\pm$ 0.022 & 0.078 & -0.57 $\pm$ 0.19 & -0.49 & 0.92 $\pm$ 0.22 &   1 &   1 & 73.03 & 1\\
K16L47V & 20.00 &  0.15 & 0.237 $\pm$ 0.115 & 0.205 & -0.50 $\pm$ 0.34 & -0.38 & 1.12 $\pm$ 0.51 &   3 &   3 & 81.05 & 1\\
K16P00N & 20.40 &  0.15 & 0.160 $\pm$ 0.068 & 0.135 & -0.31 $\pm$ 0.30 & -0.17 & 1.30 $\pm$ 0.57 &   2 &   2 & 83.46 & 1\\
K16P00T & 20.60 &  0.15 & 0.647 $\pm$ 0.050 & 0.626 & -1.61 $\pm$ 0.09 & -1.58 & 1.09 $\pm$ 0.06 &   7 &   7 & 61.80 & 1\\
K16P08O & 20.00 &  0.15 & 0.479 $\pm$ 0.172 & 0.438 & -1.11 $\pm$ 0.55 & -1.03 & 2.00 $\pm$ 1.00 &   0 &  10 & 72.99 & 0\\
K16R00W & 23.20 &  0.15 & 0.066 $\pm$ 0.020 & 0.054 & -0.67 $\pm$ 0.22 & -0.51 & 0.88 $\pm$ 0.25 &   1 &   1 & 86.70 & 1\\
K16R40M & 24.70 &  0.15 & 0.081 $\pm$ 0.012 & 0.071 & -1.49 $\pm$ 0.14 & -1.38 & 2.93 $\pm$ 0.26 &   1 &   1 & 79.68 & 1\\
K16S02G & 20.60 &  0.15 & 0.278 $\pm$ 0.092 & 0.240 & -0.88 $\pm$ 0.29 & -0.76 & 2.00 $\pm$ 1.00 &   0 &   1 & 81.21 & 0\\
K16T11B & 25.20 &  0.15 & 0.018 $\pm$ 0.006 & 0.014 & -0.34 $\pm$ 0.25 & -0.17 & 0.58 $\pm$ 0.26 &   1 &   1 & 88.54 & 1\\
K16T17X & 23.50 &  0.15 & 0.063 $\pm$ 0.017 & 0.051 & -0.75 $\pm$ 0.23 & -0.58 & 2.00 $\pm$ 1.00 &   0 &   1 & 88.14 & 0\\
K16T19Z & 23.10 &  0.15 & 0.050 $\pm$ 0.008 & 0.045 & -0.31 $\pm$ 0.10 & -0.22 & 1.14 $\pm$ 0.18 &   1 &   1 & 74.53 & 1\\
K16T56M & 26.80 &  0.15 & 0.008 $\pm$ 0.002 & 0.007 & -0.33 $\pm$ 0.21 & -0.18 & 0.54 $\pm$ 0.18 &   1 &   1 & 85.61 & 1\\
K16T57A & 20.80 &  0.15 & 0.281 $\pm$ 0.047 & 0.254 & -1.06 $\pm$ 0.21 & -0.97 & 2.80 $\pm$ 0.37 &   2 &   2 & 74.75 & 1\\
K16U25Z & 21.60 &  0.15 & 0.143 $\pm$ 0.054 & 0.131 & -0.71 $\pm$ 0.27 & -0.63 & 1.31 $\pm$ 0.45 &   1 &   1 & 72.80 & 1\\
K16U41D & 22.80 &  0.15 & 0.049 $\pm$ 0.012 & 0.044 & -0.25 $\pm$ 0.18 & -0.16 & 0.85 $\pm$ 0.23 &   2 &   3 & 74.32 & 1\\
K16UA1H & 22.70 &  0.15 & 0.067 $\pm$ 0.028 & 0.060 & -0.48 $\pm$ 0.31 & -0.39 & 1.21 $\pm$ 0.51 &   2 &   2 & 74.69 & 1\\
K16W48N & 24.80 &  0.15 & 0.144 $\pm$ 0.039 & 0.129 & -1.99 $\pm$ 0.21 & -1.89 & 2.00 $\pm$ 1.00 &   0 &   2 & 76.78 & 0\\
K16X18A & 23.10 &  0.15 & 0.065 $\pm$ 0.015 & 0.059 & -0.62 $\pm$ 0.19 & -0.53 & 2.00 $\pm$ 1.00 &   0 &   1 & 76.04 & 0\\

\label{tab.fits}
\end{longtable}
}

\end{document}